\documentclass[12pt]{template/iopart}
\eqnobysec
\usepackage{graphicx}
\usepackage{color}
\usepackage{subcaption}

\newcommand{\be}{\begin{equation}}
\newcommand{\ee}{\end{equation}}

\newcommand{\vecp}{{\mathbf p}}
\newcommand{\vecq}{{\mathbf q}}

\newcommand{\x}{{\mathbf x}}

\newcommand{\V}{{\mathbf V}}

\newcommand{\M}{{\mathbf M}}

\newcommand{\Id}{{\mathbf I_d}}

\newcommand{\der}{\partial}

\begin{document}

\title{Phase space geometry of general quantum energy transitions}

\author {Alfredo M. Ozorio de Almeida\footnote{ozorio@cbpf.br}}
\address{Centro Brasileiro de Pesquisas Fisicas,
Rua Xavier Sigaud 150, 22290-180, Rio de Janeiro, R.J., Brazil}

\begin{abstract}

The mixed density operator for coarsegrained eigenlevels of a static Hamiltonian is represented in phase space by the spectral Wigner function, which has its peak on the corresponding classical energy shell. The action of trajectory segments along the shell determine the phase of the Wigner oscillations in its interior.
The classical transitions between any pair of energy shells, driven by a general external time dependent Hamiltonian also have a smooth probability density. It is shown here that a further, contribution to the transition between the corresponding pair of coarsegrained energy levels, which oscillates with either energy, or the driving time, is determined by four trajectory segments (two in the pair of energy shells and two generated by the driving Hamiltonian) that join exactly to form a closed compound orbit.
In its turn, this sequence of segments belongs to the semiclassical expression of a compound unitary operator that 
combines four quantum evolutions: a pair generated by the static internal Hamiltonian and a pair generated by the driving Hamiltonian.
The closed compound orbits are shown to belong to continuous families, 
which are initially seeded at points where the classical flow generated by both Hamiltonians commute.

\end{abstract}

\maketitle

\section{Introduction}

The full numerical integration of the Schr\"odinger equation, required for the calculation of transitions among the energy levels of an arbitrary {\it inner Hamiltonian} $\hat H$ under the action of a general time dependent {\it driving Hamiltonian} $\hat{\Lambda}(\tau)$, becomes untenable for arbitrarily large intervals. The corresponding calculation of individual  classical trajectories of the driving Hamiltonian joining constant energy surfaces in phase space is incomparably more effective, since they are solutions of Hamilton's ordinary diferential equations. One certainly needs many trajectory segments to work out the density of classical transitions, but this amounts to parallel computation, no matter how laborious. Semiclassical (SC) approximations for energy transitions constructed on classical trajectory segments, which furnish quantum phase information supplied by trajectory actions, are thus a promissing path for the calculation of energy transitions in the hard, nonperturbative, long time regime.

SC approximations have their own difficulties, in the case of caustic singularities of the families of driven trajectories, and even for the full specification of the eigenfunctions of a nonintegrable inner Hamiltonian. In spite of this latter problem, the SC description of a coarsegrained microcanonical mixed state, combining a bunch of eigenstates within a classically narrow energy window, is quite general: it depends only on the the corresponding classical energy shell and the actions of appropriate inner trajectories that specify the phase of oscillations \cite{Ber89}. 
In a way, the classical contribution of the shell itself to this {\it spectral Wigner function} may also be attributed to short trajectories.
The duration of these segments, generated by the classical inner Hamiltonian, 
that is $t \rightarrow 0$, with zero action and hence no oscillations.
\footnote{The  time for the inner trajectories is here labled $t$, to distinguish it from a driving time $\tau$.}

The purpose of the present paper is to describe the phase space geometry  underlying oscillations in the transitions between a pair of such microcanonical states. Remarkably, the motion within each of the corresponding classical energy shells, that is, a pair of inner trajectory segments, joins on to a pair of trajectories generated by the driving Hamiltonian, so as to form a {\it closed compound orbit} (CCO). The classical action of the CCO determines the phase of its contribution to the transition density. Again, just as in the static problem, these oscillations decorate a smooth 'classical background' of zero-action CCO's, where both the inner trajectory segments are short, i.e. $t \rightarrow 0$, though the driving time $\tau$ for the pair of driving segments suffers no constraint. In other words,
the CCO's of the classical background collapse into single driven trajectories. 

The assumption of knowledge of the full unitary evolution between the initial and the final quantum states, that is the unitary evolution operator ${\hat U}(\tau)$  connecting them, allows for a considerable simplification of the present theory. This course was adopted in a sequence of papers  \cite{transI,transII,transIII}, here labeled {\bf I}, {\bf II} and {\bf III}. 
At least, one assumed knowledge of the corresponding classical canonical transformation on the phase space $\bf{R}^{2N}$ 
with its points $\{\x=(\vecp, \vecq)\}$. That is, trajectories $\tilde{\x}(\tau,\x)$ generated by the driving Hamiltonian $\Lambda(\x|\tau)$ 
(corresponding to the quantum Hamiltonian ${\hat \Lambda}(\tau)$) determine $\x\mapsto \tilde{\x}(\tau,\x) \equiv \x(\tau)$.
This necessarily preserves the energies, that is $H(\x)= H(\x(\tau))$, just as the driven quantum Hamiltonian 
\be
{\hat H}(\tau) \equiv \hat{U}(\tau){\hat H}~\hat{U}(\tau)^\dagger ~,
\label{evHam}
\ee
preserves the energy $E_k$ of each driven eigenstate $\hat{U}(\tau)|k\rangle$.

Hiding the driving Hamiltonian masks important features that depend on its interplay with the inner Hamiltonian. Even so, many of the tools developed in this sequence of papers are needed for the present general SC theory. Perforce, the general emerging qualitative picture is preserved, consisting  of a smooth classical background with a changing parameter (driving time, or one of the energies), decorated by quantum oscillations.
After all, this general scenario is not altered by our eventual knowledge of 
$\hat{U}(\tau)$. However, the assumption of full knowledge of the driving unitary transformation holds in practice only for kicked systems, such that
the motion generated by the driving classical Hamiltonian $\Lambda(\x|\tau)$ 
must be much faster than the internal motions generated by $H(\x)$. 
Otherwise, the driving trajectories must be explicitly included in the SC theory for the transition density, as here presented.  

The transition density includes a smooth classical background, which  depends only on the trajectories of the driving Hamiltonian that are responsible for the intersection of the driven shell $H(\x(\tau))=E$ with $H(\x)=E'$.
No role is played by the trajectories generated by $H(\x)$, nor by the driven Hamiltonian $H(\x(\tau))$, in this classical term.
Derived for a special case in {\bf I}, it is incorporated into the present theory, thus generalizing the result of Jarzinsky et al. \cite{Jarz} for a single degree of freedom.

In contrast, the quantum oscillations, on varying the parameters $(E,E'|\tau)$, described by the SC approximations of the transition density in the earlier treatment {\bf II} and {\bf III}, did rely on the detailed properties of compound classical orbits, each of which joins continuously an orbit segment generated by $H(\x)$ in the energy shell $H(\x)=E'$ to an orbit segment generated by the driven classical Hamiltonian 
in the energy shell $H(\x(\tau))=E$. But the integration of Hamilton's equations for such driven segments demands knowledge of the evolution 
of a neighbourhood of the entire initial energy shell $H(\x)=E$!
Unless this is given analytically, no amount of numerical integration will furnish the desired reliable driven orbits.
 
This weakness has been avoided in the present paper by replacing the previous single (unknown) driven orbit segment by a triplet of segments: a segment on the initial shell $H(\x)=E$, joined continuously to a further pair of driven trajectory segments, which transport the endpoints of the initial segment onto the $H(\x)=E'$ shell. These final points coincide with both tips of the desired driven segment, which no longer needs not be calculated in its entirety. Thus is formed a closed compound orbit (CCO) composed of
four segments instead of two, but each of these is integrated directly from the classical Hamiltonians $H(\x)$ and $\Lambda(\x|\tau)$ that are both known a priory. It will be shown in the following sections that this dismemberment of the contribution of the driven orbit segment into three parts
corresponds directly to a decomposition of the compound unitary operator, which was the basic element in the previous SC approximation of the transition density. 

The full canonical transformation driven by an external Hamiltonian is only accessible in general by integrating its individual
trajectories, but this is even more delicate if the system is time dependent. Frequently, what is here termed the (external) driving
Hamiltonian, i.e. $\Lambda(\x|\tau)$, starts out as the initial Hamiltonian itself, that is $\Lambda(\x|0)=H(\x)$. Then it is customary
to simply identify $\Lambda(\x|\tau)\equiv H(\x|\tau)$, given $H(\x|0)=H(\x)$. 
Nonetheless, this notation of the driving Hamiltonian should not be confused 
with the driven Hamiltonian $H(\x(\tau))$ corresponding to \eref{evHam}, 
since their approximate equality only arises in the adiabatic limit of very slow perturbation
of the initial Hamiltonian. For this reason, the full generality of arbitrary driving Hamiltonians is here kept as $\Lambda(\x|\tau)$,
even in the case of a weak time dependent perturbation, say
\be
\Lambda(\x|\tau) = H(\x)+ \eta f(\tau) h(\x) ~,
\label{pert}
\ee
with the small parameter $\eta$ and a corresponding expression for the quantum driving Hamiltonian.
The present theory is not constrained by any such hypothesis concerning $\Lambda(\x|\tau)$, or its quantum correspondent ${\hat \Lambda}(\tau)$.

Together with the gain in generality brought into the SC theory by an explicit role for the driving Hamiltonian, its interplay with the inner Hamiltonian will be seen to furnish the {\it seed} of small CCO's located at isolated points where the Hamiltonians initially commute. Furthermore, it will be revealed that the driving trajectory that marks the limit of a classical transition, where the driven shell of energy $E$ just touches the inner shell of energy $E'$, plays double duty as the limit of thin CCO's that contribute semiclassically to neighbouring transitions. 

The following section recalls the exact expression presented in {\bf III} of the probability density for an energy transition
as a double Fourier transform of a compound unitary operator, though this is here factored into a product of four elementary operators.
The review of the classical approximation for the transition density derived in {\bf I} is extended in section 3 to incorporate the trajectories of the driving Hamiltonian, so furnishing the condition for an {\it anticaustic} at the limit of a classical transition. The oscillatory terms that constitute the SC approximation are derived in section 4 from the SC expression of the trace of a compound product of an arbitrary number of unitary operators, 
presented in \cite{OzBro16}. Section 5 then reveals how the stationary phase evaluation of the double Fourier transform on the compound unitary operator corresponds semiclassically to a composition of Poincar\'e maps on a pair of Poincar\'e sections, together with a pair of canonical transformations between these sections. Thus each fixed point of this compound canonical transformation determines one of the relevant compound orbits. This laborious scaffolding to prove the existence of the CCO's, may then be supplanted by their continuous tracing from a seed at the point of local commutation of the pair of Hamiltonians, as presented in section 6 and further discussed in Appendix C. 

The SC approximations are developed within the context of the full Weyl-Wigner representation of quantum mechanics \cite{Groen,Moyal} 
(reviewed in \cite{Report}),
such that the density operator is represented by the Wigner function in phase space \cite{Wigner}.

\section{Energy transitions driven by general unitary operators}

Consider the evolution of pure eigenstates $|k\rangle$ of the inner Hamiltonian $\hat H$ with energy $E_k$. The one parameter family of unitary operators $\hat{U}(\tau)$ that evolve these states is generated by the driving Hamiltonian ${\hat \Lambda}(\tau)$, so that
\be
\hat{U}(\tau) = \e^{-i\tau{\hat \Lambda}/ \hbar}~
\label{drev}
\ee
if it is time independent. In the common cases where  it depends quadratically on the momentum operators, its position representation is the solution of Schr\"odinger's second order partial differential equation, even if the potential energy is time dependent. In the general case that the Hamiltonian is a general self-adjoint function of momenta and positions depending on time, the full evolution operator is defined as the time-ordered product of factors like \eref{drev} in the limit of small intervals. Such Hamiltonians can always be expressed as phase space function $\Lambda(\x|\tau)$ in the Wigner-Weyl representation, which typically differ only by SC small terms from their classical correspondents. A path integral for the evolution operator is available in this phase space representation, that is the Weyl propagator, which is completely independent of  $\Lambda(\x|\tau)$ \cite{Report}.

The probability of a transition to a state $|l\rangle$ in the driving time $\tau$ is
\begin{eqnarray}
\fl P_{kl}(\tau) = |\langle k|\hat{U}(\tau)|l\rangle|^2 = \langle k|\hat{U}(\tau)|l\rangle \langle l|\hat{U}^\dagger(\tau)|k\rangle   
={\rm tr}~\hat{U}(\tau)|l\rangle\langle l|\hat{U}(\tau)^\dagger|k\rangle\langle k| ~,
\label{Ptrans0}
\end{eqnarray}
where the projector $|k\rangle\langle k|$ is a pure state density operator.
Thus, invoking the spectral density operator \cite{Ber89}
\be
{\hat \rho}_\epsilon (E) \equiv \sum_k \delta_\epsilon(E-E_k)~ |k\rangle \langle k|
\ee
for a classically narrow energy range $\epsilon$ centred on the energy E, where 
\be
\delta_\epsilon(E)\equiv \frac{1}{\pi} ~ \frac{\epsilon}{\epsilon^2 + E^2} ~,
\label{widelta}
\ee  
one may define the probability density for the transition $E\rightarrow E'$ as
\be
\fl P_{EE'}(\tau,\epsilon) \equiv \sum_{kl} \delta_\epsilon(E'-E_k) ~ \delta_\epsilon(E-E_l) ~ P_{kl}(\tau)
={\rm tr}~\hat{U}(\tau) {\hat \rho}_\epsilon (E') \hat{U}(\tau)^\dagger(\tau) {\hat \rho}_\epsilon (E) ~.
\label{Trans}
\ee

In spite of the primordial interest in the static eigenstates and eigenenergies of the inner Hamiltonian, its associated inner unitary evolution operator
\be
\hat{V}(t) = \e^{-it{\hat H}/ \hbar}~
\label{intev}
\ee
allows for the alternative expression of the spectral density operator as
\be
\fl {\hat \rho}_\epsilon (E)
= \int_{-\infty}^\infty \frac{{\rm d}t}{2\pi\hbar}~\exp\left[ \frac{iEt}{\hbar}- \frac{\epsilon|t|}{\hbar}\right]~ \hat{V}(t) 
= \int_{-\infty}^\infty \frac{{\rm d}t}{2\pi\hbar}~\exp\left[ \frac{-iEt}{\hbar}- \frac{\epsilon|t|}{\hbar}\right]~ \hat{V}(-t) ~. 
\label{spectral3}
\ee
Inserting this in \eref{Trans} then re-expresses the transition density as the double Fourier transform
\be  
\fl P_{EE'}(\tau, \epsilon) = \int_{-\infty}^\infty \frac{{\rm d}t~{\rm d}t'}{(2\pi\hbar)^2}
 ~ \exp\left[ \frac{i}{\hbar}(E't' - Et) \right]
 ~ \exp\left[ -\frac{\epsilon}{\hbar}(|t|+|t'|)\right] ~ {\rm tr} ~ \hat{\V}(t,t'|\tau)~.
\label{PEE4}
\ee
This involves the trace of the {\it compound unitary operator}
\be
\hat{\V}(t,t'|\tau) \equiv  
\hat{U}(\tau) ~ \hat{V}(t')~\hat{U}(-\tau) ~ \hat{V}(-t),
\label{compuni}
\ee
now dismembered as a product of four operators, 
instead of just two in {\bf III}. In other words, the definition of the {\it driven inner evolution operator}
\be
\hat{V}(t'|\tau) \equiv \hat{U}(\tau) ~ \hat{V}(t')~\hat{U}(\tau)^\dagger,
\label{compact} 
\ee
reduced \eref{compuni} into a product of a pair of unitary operators. Indeed, the pair of $U$-operators acting on the $V$-operators are an example of the {\it superoperators} sudied semiclassically in \cite{SarOA16}.

\section{Review of the classical approximation}

Corresponding to the discrete energy spectrum, that has here been assumed,
one presumes a bound classical system, so that the phase space volume of the $(2N-1)$-D energy shells $H(\x(\tau))=E$ increases monotonically with $E$.
The classical approximation for the transition density depends on the driven  shell, together with the initial $H(\x)=E'$ shell, as derived in {\bf I}
in the special case that $\hat{U}(\tau)$ is a phase space reflection operator \cite{Report}. 
Indeed the intersection of the driven energy shell 
$H(\x(\tau))=E$ on $H(\x)=E'$ determines a surface with dimension $2(N-1)$, which may be considered to define a generalized Poincar\'e section of this stationary shell, named the $E'$-section. This reduces to a pair of points for $N=1$, the case treated in \cite{Jarz}. 
Reversing this procedure, one may now drive the Hamiltonian backwards to obtain the pre-image of this shell, that is, $H(\x(-\tau))=E'$, which, in its turn, defines a further $2(N-1)$-D section, but this time
on the original $H(\x)=E$ shell, the $E$-section.

The spectral density operator for the driven energy shell is the Fourier transform \eref{spectral3} of the driven inner evolution operator, 
which implies that
\be
{\hat \rho}_\epsilon (E'|\tau ) =  
\hat{U}(\tau) ~{\hat \rho}_\epsilon (E') ~\hat{U}(\tau)^\dagger ~.
\ee
Let us turn now to the Wigner-Weyl representation, in which the spectral operators are represented by spectral Wigner functions. Then the simplest classical approximation for the spectral Wigner functions are \cite{Ber89,Ber89b}
\be
W_{\epsilon}(\x,E') \approx  \delta(H(\x) - E') ~~~{\rm and} ~~~ 
W_{\epsilon}(\x,E|\tau) \approx  \delta(H(\x(\tau)) - E) ~,
\ee
and recalling that the trace of a product of operators is rendered by the integral over the product of Wigner functions \cite{Report},
one obtains alternative, but equivalent expressions for the classical transition density:
\begin{eqnarray}
P_{EE'}(\tau,\epsilon)_{00}& \approx \frac{1}{(2\pi\hbar)^{N}}\int d\x ~ \delta(H(\x) - E') ~ \delta(H(\x(\tau)) - E) \\   \nonumber
&= \frac{1}{(2\pi\hbar)^{N}}\int d\x ~ \delta(H(\x(-\tau)) - E') ~ \delta(H(\x) - E) ~.
\label{Ptrans2}
\end{eqnarray}
In other words, one either integrates a Dirac delta-function over the (forwards) $E'$-section, or over the (backwards) $E$-section,
as shown in Fig. 1(a).
Thus, following {\bf I}, one obtains the generalized classical contribution as
\be
\fl P_{EE'}(\tau,\epsilon)_{00} \approx \oint_{E'_s} d\x ~|\{H(\x),H(\x(\tau))\}|^{-1}
= \oint_{E_s} d\x ~|\{H(\x),H(\x(-\tau))\}|^{-1} ~,
\label{Ptrans7}
\ee
where the single Poisson bracket here matches the Hamiltonian with its forward evolved image on the $E'$-section ($E'_s$),
or its backward image on the $E$-section ($E_s$).

\begin{figure}
\centering
\begin{subfigure}{.5\textwidth}
  \centering
  \includegraphics[width=.8\linewidth]{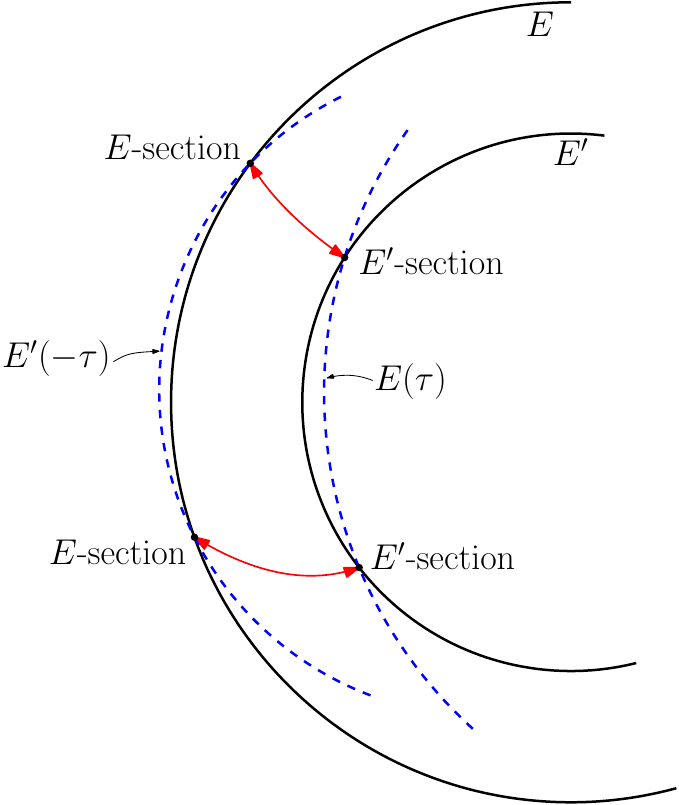}
  \caption{}
\end{subfigure}%
\begin{subfigure}{.5\textwidth}
  \centering
  \includegraphics[width=.7\linewidth]{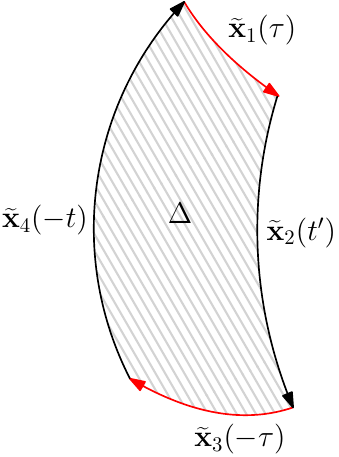}
  \caption{}
  \label{fig:sub2}
\end{subfigure}
\caption{(a) The classical Hamiltonian $\Lambda(\x|\tau)$ drives the energy shell $H(\x)=E$ so that it intersects the $H(\x)=E'$, forming the $E'$-section in the time $\tau$. Conversely, it drives the $H(\x)=E'$ to intersect the $H(\x)=E$ shell in the time $-\tau$, forming the $E$-section. Both these sections are $2(N-1)$-D surfaces, which reduce to a pair of points for $N=1$.
(b) The compound trajectory is formed by four segments. Denoting each driven segment simply by $\tilde{\x}_j(\tau)$, the pair of segments
$\tilde{\x}_1(\tau)$ and $\tilde{\x}_3(-\tau)$ start on the $E$-section
and the $E'$-section respectively, both driven by the Hamiltonian $\Lambda(\x|\tau)$, whereas $\tilde{\x}_2(t)$
and $\tilde{\x}_1(-t)$ are generated by $H(\x)$. The latter trajectory segments induce the Poincar\'e maps on the $E'$-section
and the $E$-section respectively. Note that for $N=1$ there is no distinction between the trajectory and the energy shell, 
so the same figure is used for (a) and (b).}
\label{fig:test}
\end{figure}

In spite of the dimensional reduction in this last formula, one should note that it already introduces in the classical approximation
the obstacle that one wishes here to avoid. The Poisson bracket in the integrand requires the knowledge of the derivatives of both
Hamiltonians along the appropriate section. However, even if the initial $H(\x)=E'$ shell is given analytically, 
the driven $H(\x(\tau))=E$ shell generally requires a numerical computation. Thus, it would require a huge numerical effort 
to compute reliably the Poisson bracket on the entire $E'$-section. On the other hand, a Montecarlo estimation of the classical
transition density directly via (3.4) is quite feasible by present numerical methods. 

In the following SC approximation, which adds oscillations of the transition density to the smooth classical background, 
unknown trajectories, generated by Hamilton's equations for the driven system, are no longer employed. 
On the other hand, the Poincar\'e map for the trajectory segments generated by the initial Hamiltonian $H(\x)$, 
on both its $E'$-section and its $E$-section,
define segments of the compound orbits, which are the classical basis for these oscillations. One should note that each of these sections
is merely indicated in Fig 1a by a pair of points.

Of course, the very existence of these component orbit segments depends on the occurrence of an intersection of a driven energy shell with its static partner.
For a finite energy difference it takes a finite time: $\tau_m$ for them to touch, before they can intersect transversely. But at this instant
\be
\{H(\x),H(\x(-\tau_m))\} = 0 ~,
\label{anticaustic}
\ee
so that the integrand in (3.4) is singular. On the other hand, 
the intersection of the energy shells reduces to a point, 
so that there are conflicting limits. The analysis in {\bf I} for a pair of convex energy shells concludes that the integral converges for $N>1$. Until further possibilities for touching energy shells are examined, this result will be assumed.  

So it is only in the case of a single degree of freedom that a full fledged caustic has to be unraveled by an improved SC approximation for the spectral Wigner function as presented in {\bf I}. This dependence on the phase space dimension resembles that of the projection of spectral Wigner functions into the corresponding mixed wave functions. There it was found by Berry \cite{Ber89b} that only when $N=1$ does a real caustic singularity arise, which requires a higher Airy function approximation. Otherwise, a mild {\it anticaustic} bounds the region where the wave function, or here the transition density, is appreciable. Outside, the classical approximations merely neglect exponentially small terms. 

This nomenclature will be incorporated here, that is, fixing the transition energies $E$ and $E'$, the point $\x_a$, where a driven shell touches a static inner shell after a time $\tau_a$, will be referred to as the anticaustic, since we are basically concerned with systems where $N>1$. This holds for the initial contact $\tau_m=\tau_a$, or after a full traversal of the driven $E$-shell through the $E'$-shell, which may occur for very large $\tau_a$, or indeed any intermediate nontransverse intersection of the shells. 
\footnote{Notice that\eref{anticaustic} implies that $H(\x)$ is constant along the direction of the velocity generated by $H(\x|\tau)$ and vice versa. 
Both these vectors lie in the tangent planes of the energy shells of the respective Hamiltonians, which are in this way constrained.} 
Generally, the $E'$-section shrinks to the point $\x_a$ as $\tau \rightarrow \tau_a$. This is represented by the coalescing pair of points on the $E'$-shell in Figure 1(a),
which becomes a shrinking $2(N-1)$-D disk as discussed in {\bf I} and {\bf II}.
Then $\x_a$ is the end point of a single trajectory segment joining the two shells, that is, $\tilde{\x}_1(\tau_a) = \tilde{\x}_3(\tau_a)$ in Figure 1(b).

Initially, it is only the 'diagonal transition', i.e. for energy preservation, that has a super classical singularity, because the driven shell fully coincides with the static shell. However, at this limit the probability of energy preservation is one, so one will not attempt its classical or semiclassical unraveling.

In terms of the driven trajectories $\tilde{\x}(\x,\tau)$ generated by 
$\Lambda(\x|\tau)$, the $E'$-section is given by the segments with endpoints satisfying
\be
H(\x) = E'  ~~~~ {\rm and} ~~~ H(\tilde{\x}(\x,-\tau))=E ~,
\label{drivseg}
\ee
while the $E$-section, also shown in  Figure 1(a), is defined by the time inverse of these segments. Thus, each of these sections is defined by the Hamiltonian flow from the other as a one-to-one map, since Hamilton's equations are of first order.

The definition of the {\it stability matrix} for each driven trajectory segment as $\Gamma(\x|\tau)$, which is responsible for the linearized canonical transformation in the  trajectory neighbourhood, simply transports the inner velocity vector
$\dot{\x}(\x)$ in the $E$-shell to the driven velocity vector in the driven shell:
\be
\dot{\x}(\tilde{\x}(\x,\tau)|\tau) = \Gamma(\x|\tau) ~ \dot{\x}(\x).
\ee 
Therefore, the Poisson bracket in \eref{Ptrans7} and \eref{anticaustic} can be directly expressed in terms of the driven trajectory segments \eref{drivseg} together with their respective stability matrices, that is,
\be
\fl \{H(\x),H(\x(\tau))\} 
= \dot{\x}(\tilde{\x}(\x,\tau)) \wedge \dot{\x}(\tilde{\x}(\x,\tau)|\tau)
= \dot{\x}(\tilde{\x}(\x,\tau)) \wedge [\Gamma(\x|\tau)~ \dot{\x}(\x)] ~.
\ee 

As an anticaustic is approached, that is, $\tau \rightarrow \tau_a$, 
the shrinking of the $E'$-section compresses the corresponding driven trajectory segments into a narrow tube, finally reducing to a single segment,
which has  $\x_a$ as its endpoint. However, at $\x_a$ the driven $E$-shell must be tangent to the $E'$-shell, that is,
\be
\frac{\der H}{\der\x}(\x_a) = \frac{\der H}{\der\x}(\tilde{\x}(\x_a(-\tau)) ~,
\ee 
or, equivalently
\be
\dot{\x}(\x_a) 
= \Gamma((\tilde{\x}(\x_a,-\tau))|\tau)~ \dot{\x}(\tilde{\x}(\x_a,-\tau)) ~.
\label{anticaustic2}
\ee

It is then important to realize that this set of $2N$ equations on $2N$ phase space coordinates does generally define an isolated point, but it is much stronger than the condition for a singularity of the integrand of \eref{Ptrans7}, which may be re-expressed
as
\be
\dot{\x}(\x_a) \wedge [\Gamma((\tilde{\x}(\x_a,-\tau))|\tau)~ \dot{\x}(\tilde{\x}(\x_a,-\tau))] = 0 ~.
\label{anticaustic3}
\ee 

Hence, \eref{Ptrans7} has a singular integrand along a codimension-1 manifold, that is, $(2N-1)$ dimensions in the full phase space, intersecting the $E'$-section in a $(2N-3)$-manifold. However, it depends on the inner velocity fields at the tips of a driven trajectory segment. It may not be present in a given section and, as stated previously, the singularity should be integrable if $N>1$.

\section{Semiclassical approximations}

The stationary phase evaluation of the integral \eref{PEE4} for the transition density, 
over the trace of the compound unitary operator \eref{compuni}, depends on its SC description in terms of classical trajectories.
The original discussion of compound evolution operators in the context of evolving quantum correlations
considered products of an arbitrary number of evolutions in the Wigner-Weyl representation. It turns out that the SC trace of a general compound operator derived in \cite{OzBro16} depends on the action of general CCO's: they join trajectory segments from each of the evolutions into piecewise smooth
closed curves in phase space. The previous constructions in {\bf II} and {\bf III} concerned just a pair of curves, one evolved by
the classical Hamiltonian $H(\x)$ in the positive or negative time $t$ and the other evolved by the driven Hamiltonian $H(\x(\tau))$
in the time $t'$. Now, the unfolding in \eref{compuni} of the driven unitary operator \eref{compact}
into its three components is reflected in the replacement of the driven classical segment by three curves:
the segment in the evolved shell $E(\tau)$ in Fig 1a is replaced by the segments $\tilde{x}_1, \tilde{x}_2, \tilde{x}_3$
shown in Fig. 1b. Therefore the expanded compound trajectory contains no segment generated by a Hamiltonian that is not known ab initio.
\footnote{This formation of a curved tetragon, describing the evolution of one of its sides, recalls the SC description of superoperators \cite{SarOA16}.}

Thus, one defines $\S(t, t'|\tau)$, the classical action of the CCO, that is, the total action generated by the piecewise continuous Hamiltonian: $\Lambda(\x| \cdot)$ from zero to $\tau$, followed by $H(\x)$ from zero to $t'$,
then $\Lambda(\x| \cdot)$ drives from zero to $-\tau$ and finally $H(\x)$ again from zero to $-t$. Therefore, $\S(t, t'|\tau)$ is the total
piecewise continuous time-dependent action, for the full CCO 
$(\tilde{\x}_1(\cdot), \tilde{\x}_2(\cdot),\tilde{\x}_3(\cdot), \tilde{\x}_4(\cdot))$ with period $\tau+t'-\tau-t$, shown in Fig 1b.
Of course, it should be recalled that the 2-D illustrations play double duty for the $(2N-1)$-D energy shells and for the relevant orbits
which they contain. In this way, the pair of trajectory segments $\tilde{\x}_1$ and $\tilde{\x}_3$ belong to the set that generate the $E$-section and the $E'$-section sketched in Figure 1a. 

According to \cite{OzBro16}, the SC contribution of each CCO to the trace is
\begin{equation}
{\rm tr } ~ \hat{\V}(t, t'\tau) \approx \frac{2^N}{|\det[\Id - \M(t, t'|\tau)]|^{1/2}}\>\>
\exp \left[ \frac{i}{\hbar}(\S(t, t'|\tau)+ \hbar \sigma)\right] ~.
\label{Uweyl}
\end{equation}
The linear approximation of the transformations generated by $H(\x)$ near the $t'$-trajectory and the $-t$ trajectory are defined 
by the respective {\it stability matrices} $\M(t')$ and $\M(-t)$, just as the stability matrices $\bf{\Gamma}(\tau)$ and $\bf{\Gamma}(-\tau)$ account for the neighbourhood of the segments driven by the classical Hamiltonian $\Lambda(\x|\tau)$. Then the full stability matrix for the compound trajectory is simply
\be
\M(t, t'|\tau) \equiv \M(-t)\bf{\Gamma}(-\tau)\M(t') \bf{\Gamma}(\tau)~, 
\label{stabcomp}
\ee
whereas $\Id$ is the identity matrix. 

There is a strong resemblance of \eref{Uweyl} with the Gutzwiller trace formula for the (static) density of energy levels \cite{Gutzbook}. However,
the single periodic orbit there already appears with an energy action, after the Fourier transform over time. Also, one should note that the discontinuity of the derivative of the compound trajectory
at the initial and final points of each trajectory segment implies that generally there is no zero eigenvalue of the
full stability matrix, unlike smooth periodic trajectories. On the other hand, the determinant in the amplitude of \eref{Uweyl}
does not depend on the choice of the starting point of the CCO.   
 
There may be multiple branches of the action, meeting along caustics in the $(t,t'|\tau)$ space. These denote bifurcations of the families of CCO's, where the semiclassical amplitude diverges,
that is, $[\det\Id - \M(t,t'|\tau)]=0$.
The phase $\sigma$ is determined by the convergence of neighbouring paths.
(For convenience of notation, all such focal indices shall henceforth be denoted by the generic symbol $\sigma$, without specifying 
their values, which are altered by the Fourier transformations carried out.)
Generally there is an increment on $\sigma$ where a caustic is crossed.

Fixing the evolution parameter $\tau$, the evaluation of the integral over $t$ and $t'$ in \eref{PEE4} by stationary phase requires the time derivatives of $\S(t, t'|\tau)$. Clearly, this derivative for the action of either separate segment, generated by the initial Hamiltonian $H(\x)$, is just the respective energy. It is shown in Appendix A that this also holds for the CCO, so that 
\be
\frac{\der \S}{\der t}(t, t'|\tau) = -t~E(t) ~~~~  {\rm and} ~~~~ 
\frac{\der \S}{\der t'}(t, t'|\tau) = -t' ~E(t') ~,
\label{stat}
\ee
where $E(t)$ and $E(t')$ are the respective energies of the $t$-segment and the $t'$-segment. Therefore the stationary phase conditions for the integrand \eref{Uweyl}, within the double integral in \eref{PEE4}, 
are simply $E(t)=E$ and $E(t') = E'$.  

The double stationary phase condition for \eref{Uweyl}, including the trajectory segments driven by $\Lambda(\x|\tau)$, $\tilde{\x}_j(\tau')$ and $0<\tau'<\tau$, places the trajectories $\tilde{\x}_4$ and $\tilde{\x}_2$
on the the energy shells with energy $E$ and $E'$, that is, the  $E$-segment with the $E'$-segment, 
forming the curvilinear quadrilateral in Fig. 1b with symplectic area $\Delta(E,E'|\tau)$. Thus, the resulting total stationary action for the final CCO can be expressed as
\be
\S(E, E'|\tau) = \Delta(E,E'|\tau) 
- \int_0^\tau \Lambda(\tilde{\x}_1(\tau')|\tau')~ d\tau' - \int_0^{-\tau} \Lambda(\tilde{\x}_3(\tau')|\tau')~ d\tau'.
\label{totact}
\ee
 
Finally, combining the complex conjugate contributions of the compound orbits in \eref{PEE4}
with their time reversal, the SC transition density takes the the form:
\begin{eqnarray}
P_{EE'}(\tau, \epsilon) \approx & \frac{2^N}{\pi\hbar} \sum_{j,j'} 
\exp\left( -\frac{\epsilon}{\hbar}(|t_j|+|t'_{j'}|)\right)
\left|\det \frac{\der(t_j,t'_{j'})}{\der(E,E')} ~ \right|^{1/2} 
\\  \nonumber
& |\det[\Id - \M(t_j, t'_{j'}|\tau)]|^{-1/2} ~\cos \left[ \frac{1}{\hbar}\S_{j,j'}(E, E'|\tau)+ \sigma\right] ~.
\label{PSC}
\end{eqnarray}

It will be shown in the followig section how each resulting CCO, for fixed $(E,E'|\tau)$, determines a fixed point of the compound canonical transformation correponding to the Fourier transform of the compound unitary operator $\hat{\V}(t,t'|\tau)$.
Therefore, CCO's arise in continuous families $t_j(E,E'|\tau)$ and $t_{j'}(E,E'|\tau)$, each with its action $\S_{j,j'}(E, E'|\tau)$, and hence they determine contributions to the transition density, which describe SC oscillations with variation of any of the parameters.

In {\bf II} and {\bf III} the action resulting from the stationary phase integration reduced to the symplectic area of a simpler two-component CCO, which was purely energy dependent. Indeed, the $\tau$-dependence of the SC contributions to the transition density analogous to (4.5) was hidden within the time dependence of the $\tau$-driven $E$-segment.

 \section{Compound maps joining the Poincar\'e sections}

 The SC description of energy transition densities that evolve with time depends on families of CCO's, which, though purely classical, are not relevant in any familiar context in classical dynamics. It is then necessary to establish their very existence, their family relations and, if possible, how to find them. So, one starts by detecting them within appropriate generalized Poincar\'e sections.    

The fact that the Poincar\'e maps generated by $H(\x)$ on both the $E$-section and the $E'$-section are canonical 
follows from the the Poincar\'e-Cartan theorem \cite{Arnold, livro}, as it guarantees the preservation of
\be
\oint\left[ \vecp \cdot d\vecq - H(\x) dt' \right] ~
\ee
in the extended $(2n+1)$-D extended phase space, which reduces to the symplectic area itself on both the $E$-section and the $E'$-section.
But this property also holds for the mapping between them that is generated by the driving Hamiltonian $\Lambda(\x|\tau)$ 
for a fixed duration. Indeed, the Poinar\'e-Cartan invariant  
\be
\oint\left[ \vecp \cdot d\vecq - \Lambda(\x|\tau') d\tau' \right]
\label{PCinvariant}
\ee
on any tube of trajectories also reduces to its symplectic area at the intersection of the tube with the initial section and
on its intersection with the final section at $\pm\tau$.

Thus these four canonical transformations can now be sequenced into a canonical compound map, 
in analogy with the compound unitary operator \eref{compuni}.
This is not a strict correspondence, because the operator is defined on the full Hilbert space, whereas the present compound mapping
is based on the pair of $2(N-1)$-D Poincar\'e sections defined in the previous section. 
Indeed, the endpoints of each of the four trajectory segments portrayed in Fig. 1b are either mapped into each each other 
by a Poincar\'e map by $\tilde{\x}_2$ internal to the $E'$-section, or to the $E$-section by $\tilde{\x}_4$. 
Interposed between these, an equal $\tau$-time canonical map of trajectory segments $\tilde{\x}_1$ takes the $E$-section to the $E'$-section, 
while $\tilde{\x}_3$ reverses this map. 
Each piecewise continuous sequence of four trajectory segments is a CCO in the full phase space,
which determines a fixed point of the compound canonical map. Conversely, the fixed points of the compound map determine 
the desired closed compound orbits, which support the stationary phase solutions in the previous section.
\footnote{The previous nomenclature, which distinguished {\it orbits} in an energy shell from time-dependent {\it trajectories}, 
is no longer tenable for the present hybrid compound orbits.}  

It should be recalled that, just as in {\bf II} and {\bf III}, the Poincar\'e mappings are here defined by odd traversals
of the section, instead of the more familiar case of even traversals. So, for the case of the $(2j-1)$-traversal of the $E$-section
and the $(2j'-1)$-traversal of the $E'$-section, a fixed point of the compound mapping defines the $jj'$ compound orbit.
\footnote{So as not to encumber the notation, the sums over $j$ and $j'$ include all the compound orbits for each pair of traversals: the fixed points of the full compound map for the corresponding winding of the factor Poincar\'e maps.}
Each of the stationary phase approximations of \eref{PEE4} combines a pair of trajectory segments generated by the driving Hamiltonian 
$\Lambda(\x|\tau)$ with a pair of orbit segments in the chosen energy shells of $H(\x)$ into a piecewise smooth curvilinear quadrilateral.
The continous $jj'$-families of $(E,E'|\tau)$-dependent compound orbits are in one-to-one relation with each of the fixed points 
of the continuous $(E,E'|\tau)$-dependent families of $jj'$-compound maps. Likewise, the periods for the orbit segments 
in each Poincar\'e map are smooth functions $t_j(E,E'|\tau)$ and $t_j'(E,E'|\tau)$. Then, even though the amplitude 
of each stationary phase contribution is the determinant of the Hessian matrix of the total time dependent action \eref{totact},
the final expression for the oscillatory terms of the transition density is already expressed in (4.5) in terms of
\be
\frac{\der^2}{\der t\der t'}\S(t,t'|\tau) = \det \frac{\der(E_j,E_{j'})}{\der(t,t')} 
= \left[\det \frac{\der(t_j,t'_{j'})}{\der(E,E')} \right]^{-1}.
\label{Jacinv}
\ee
The {jj'}-sum in (4.5) includes all combinations of forward and backward trajectory segments, 
always obtained as fixed points of the corresponding
compound map, but of course very long orbits will be exponentially dampened by the factors depending on the 
energy width $\epsilon$. One must add to this the classical contributions of 'zero-length' orbits (the section itself) 
presented in section 3, to obtain the complete portrayal of the energy transition density.

\section{Seeds for families of closed compound orbits}

The one-to-one relation of the trajectories generated by a single Hamiltonian to those of either a Poincar\'e map or a map with fixed time identifies continuous periodic orbits with fixed points that are generally isolated. Parameterized by either time or energy, fixed points form smooth families, except at bifurcations. The analogous identification of each CCO with the fixed point of a compound map also guarantees its insertion in a continuous family parameterized by either $(E,E'|\tau)$ or $(t,t'|\tau)$. Thus, once a single fixed point of the compound map has been found and the full CCO drawn out, one may generate the family by repeated use of a generalized Newton's method \cite{Bar,Aguiar}.

In spite of this considerable simplification, the search even for a single CCO with fixed parameters remains elusive, since these structures are not relevant in a purely classical context. One should not need to propagate a pair of energy shells, so as to determine a pair of generalized Poincar\'e sections, only then to determine the fixed points of the composition of four canonical maps! 

So, constraining the internal times $t=t'$, let us first specify a CCO in an alternative fashion by reversing the respective times in segments $\tilde{\x}_3$ and $\tilde{\x}_4$ of Figure 1(b).
Then a CCO exists if, for a given initial point, both the piecewise smooth trajectories $[\tilde{\x}_1(\tau),\tilde{\x}_2(t)]$ and $[\tilde{\x}_4(t),\tilde{\x}_3(\tau)]$ share the same endpoint. But this is tantamount to requiring that, for this initial value, the internal motion  generated by $H(\x)$ commutes with the driven motion. 

Consider now the limit where both $t$ and $\tau$ are negligible. It is easier to begin with a time independent driving Hamiltonian $\Lambda(\x)$. Then the trajectory segments composing the CCO can be identified with the Hamiltonian vector fields, $\dot{\x}_H(\x)$ and $\dot{\x}_\Lambda(\x)$ generated by $H(\x)$ and $\Lambda(\x)$, so that the total time derivative of any phase space function  $\phi(\x)$ along either flow is the action of the differential operators
\be
\frac{d}{dt}= \dot{\x}_H \cdot \frac{\der}{\der \x} ~~ and~~~
\frac{d}{d\tau}= \dot{\x}_\Lambda \cdot \frac{\der}{\der \x}  ~.
\ee
The difference in the increment of $\phi(\x)$ due to a pair of short propagations by $\delta t$ and $\delta\tau$, on exchanging their order, is then defined by 
\be
\delta \phi(\x) \approx 
\delta t ~ \delta \tau \left[ \frac{d}{dt}, \frac{d}{d\tau} \right] \phi(\x)
= \delta t \delta \tau \left( \frac{d}{dt}\frac{d}{d\tau} - \frac{d}{d\tau} \frac{d}{dt}\right) \phi(\x) ~,
\ee
introducing the {\it differential commutator} in the second term. Hence, a differential CCO is defined at any phase space point $\x_0$, where
\be
\left[ \frac{d}{dt}, \frac{d}{d\tau} \right]_{\x_0}=0 ~.
\ee

Furthermore, according to Arnold in chapter 8 of \cite{Arnold}, the differential commutator coincides with the time derivative along the vector field that is generated by $\{H(\x), \Lambda(\x)\}$ (i.e., the Poisson bracket of the internal Hamiltonian with the driving Hamiltonian) acting as a Hamiltonian in its own right.
In fact, one may remove the restriction on time dependence of the driving Hamiltonian, so as to state that the {\it seed} of a family of CCO's is thus an equilibrium point $\x_0$ of the {\it Poisson bracket Hamiltonian} 
$\{H(\x_0), \Lambda(\x_0|\tau=0)\}$, that is,
\be
\frac{\der} {\der \x} \{H(\x), \Lambda(\x|\tau=0)\} |_{\x_0}= 0 ~.
\ee
On it is formed the differential CCO, a parallelogram of commuting vectors generated by $H(\x)$ and $\Lambda(\x|0)$ at $\x_0$, multiplied by the small times $\delta t$ and $\delta \tau$. From this point, one may follow the family for increasing $\tau$, or $t$, as well as taking $t\neq t'$. The initial value for these commuting trajectory segments will also wander off from the seed at $\x_0$. One should note that starting from the seed, it makes no difference that the instantaneous vector field of the driving Hamiltonian varies in time, because we only need it initially to reveal the seed. The existence of the family for longer times is guaranteed by the continuous existence of a fixed point of the compound map, which needs only invoked, since it is found directly by following the family of CCO's.

A simple example, just to fix ideas, is a simple harmonic oscillator
\be
H(\x) = \frac{p^2}{2} + a \frac{q^2}{2} ~,
\ee 
driven by a displaced oscillator
\be
\Lambda(\x) = a \frac{p^2}{2} +  \frac{(q-b)^2}{2} ~,
\ee 
so that the Poisson bracket Hamiltonian is
\be 
C(\x)= \{H(\x), \Lambda(\x)\} = (1-a^2) pq - bp.
\ee 
The seed then lies at the zero of its velocity field
\be
\frac{\der C}{\der p} = (1-a^2)q -b = 0 ~~~{\rm and} ~~~
\frac{\der C}{\der q} = (1-a^2)p = 0, 
\ee 
furnishing the seed at
\footnote{Curiously, this is an unstable equilibrium of $C(\x)$ and this is also the case if $\Lambda(\x)$ is instead an inverted oscillator.}
\be
p_0 = 0 ~~~{\rm and} ~~~ q_0 = \frac{b}{1-a^2} ~.
\ee 
For higher than quadratic Hamiltonians there may be more equilibria seeding CCO families.

The seed of a family of CCO's is generically located at an isolated phase space point $\x_0$, which contributes an initial CCO, i.e. at $\tau=0$, related to the preserved energy $E_0 = H(\x_0)$. Actually, any energy must be preserved as 
$\tau \rightarrow 0$, so one should seek CCO's satisfying this limit at other energies. Indeed, what is required is an inner trajectory segment of time $t$ for which the driven velocities $\dot{\x}_\Lambda(\x)$ at both ends are mapped into each other by the stability matrix $\M(t)$ of the inner motion, that is, for an initial value as $\x$,
\be
{\dot{\x}}_\Lambda(\tilde{\x}(\x,t)) = \M(\x,t) ~{\dot{\x}}_\Lambda(\x)~.
\label{CCOt}
\ee 
For each inner time $t$, this is a set of $2N$ equations for the $2N$ initial values $\x(t)$, which can be solved by iterated use of Newton's method, for growing $t$, starting from the seed at $\x(t) \rightarrow \x_0$ as $t\rightarrow 0$. The energy of each trajectory segment is 
$E(t) = H(\x(t))$, so that this continuous family of segments in a continuous range of energies forms the edge of the family, a t-subfamily of thin CCO's,  which broaden as the driving time $\tau$ increases.

Each initial thin CCO can only contribute to a diagonal term for the energy transition, that is, to the probability of no transition. But this is unity  anyway at $\tau=0$ and the precise way that this limit is achieved semiclassically is not of present concern. For finite $\tau$, the pairing inner orbit ($\tilde{\x}_4$ to $\tilde{\x}_2$ in Figure 1(b)) need not have the same energy and indeed, we may relax the constraint that $t=t'$ and, hence, run through a range of initial and final energies. 

A further comment is due concerning the common form of perturbative Hamiltonian \eref{pert}. The
condition for the seed then ignores the stationary part of the Hamiltonian, that is, one only requires
\be
\frac{\der} {\der \x} \{H(\x), h(\x)\} |_{\x_0}= 0 ~.
\ee
Even though this defines the same seed as for the stationary Hamiltonian $h(\x)$ on its own, the trajectories that evolve from a finite initial segment will be quite distinct: the ones generated by $h(\x)$ will generally depart transversely from its energy shell, whereas the pair generated by the full Hamiltonian $\Lambda(\x|\tau)$ will spiral slowly away. This does not alter the topology for their connecting to another segment driven by $H(\x)$
on another energy shell.

Consider now the other limit of thin CCO's with finite driving time $\tau$, 
but $t=t' \rightarrow 0$. In this limit, one obtains again a single segment, but now it is the corresponding stability matrix
$\Gamma(\x|\tau)$ that needs to map the initial inner velocity 
${\dot{\x}}_H(\x)$ to the other end of the segment, that is,
\be
{\dot{\x}}_H(\tilde{\x}(\x,\tau)) = \Gamma(\x|\tau) ~{\dot{\x}}_H(\x)~.
\label{CCOtau}
\ee
This $\tau$-subfamily of thin CCO's can be grown continuously from the seed at
$\x_0$, just as the $t$-subfamily specified by \eref{CCOt}, so that it forms the other edge of the full three parameter family of CCO's. Each
segment in the $\tau$-subfamily $\tilde{\x}(\x,\tau)$ maps 
$E(\tau)=H(\x) \mapsto E'(\tau)=H(\tilde{\x}(\x,\tau))$ and thus supplies a zero action SC contribution to the transition density at this pair of energies.

We can now identify each endpoint of a thin segment in the $\tau$-subfamily with an anticaustic. Indeed, \eref{CCOtau} is merely the time reverse of \eref{anticaustic2}.
Thus, the compression of all the trajectory segments, which form both the $E'$-section and the $E$-section, as $\tau \rightarrow \tau_a$ brings within it the compression of a CCO, such that $|\tilde{\x}_4 \rightarrow \tilde{\x}_2| \rightarrow 0$ in Figure 1(b). 
Then the action of a CCO in the $\tau$-subfamily has zero action and so has the phase of its contribution to the transition density. 
On the other hand, the amplitude of its SC contribution is singular, because in this limit the compound stability matrix \eref{stabcomp} reduces to
\be
\M(t, t'|\tau) \rightarrow 
\mathbf{I}_d ~\bf{\Gamma}(-\tau_a)~\mathbf{I}_d ~\bf{\Gamma}(\tau_a) =\mathbf{I}_d~. 
\label{stabcomp2}
\ee
The simple SC contribution in the neighbourhood of a thin CCO should then be replaced by an improved uniform approximation to avoid its spurious singularity, such as supplied in {\bf I} for the caustic as two shells touch if $N=1$. Still, the simple SC approximation indicates the presence of a peak of the CCO contribution, which may turn out to dominate that of the classical anticaustic. Future work is thus needed to establish the full structure of the transition density in a small neighbourhood of the anticaustics. 

There may still linger a doubt that a few seeded families of CCO's can account for all the $jj'$-families CCO's that should be signaled by fixed points of the energy dependent compound map, for each new winding of its factor Poincar\'e maps. 
However, starting with any segment in the thin $\tau$-family, there is no limit to the increase of either inner time, $t$ or $t'$ for fattening CCO's.
The respective segments will wind as many times as required along the pair of energy shells with energy $E(t,t'|\tau)$ and $E'(t,t'|\tau)$, though the contributions of these CCO's to the transition density is expoentially dampened with increasing $t$ or $t'$.

\section{Discussion}

Determining the fixed points of compound maps for each value of the parameters $(E, E'|\tau)$, together with the actions and the stability of the corresponding compound orbits, would certainly be a demanding computational task, even though the finite energy window effectively cuts off full orbits with $\epsilon(|t_j|+|t_{j'}|) >>\hbar$. 
Fortunately this procedure can be entirely avoided by growing directly
each full family of CCO's from its seed as functions of $(t,t'|\tau)$ and then evaluating the dependent variables $E(t,t'|\tau)$ and $E'(t,t'|\tau)$. 
This follows from a simple adaptation of the method 
developed by Michel Baranger and coworkers \cite{Bar,Aguiar}, for following through continuous bifurcation trees of periodic orbits. All that is needed is to feed in a good discretization of the orbit at the previous parameter to a multidimensional Newton's method, so as to quickly converge onto the closed compound orbit at the next parameter.
In this way, it is possible to determine nearly continuous sequences of compound orbits, bypassing the need to construct the corresponding
compound maps and then search for their fixed points at each step.  

The special case of of compound orbits, in which one of the segments lies on a periodic orbit, was singled out in {\bf II}, because iterations
of the relevant Poincar\'e map bring in further contributions with a phase difference proportional to the periodic orbit action. Clearly,
there is no distinction with respect to the present scenario for a periodic orbit in the initial $E'$-shell, that is, $\tilde{\x}_2$ in Fig 1b.
But now the other segment ($\tilde{\x}_4$) lies in the initial $E$-shell, 
instead of the driven segment $\tilde{\x}_4(\tau)$ on the driven $E$-shell. All the same, being canonically related, 
both these segments belong to periodic orbits with the same action, so that all the discussion about possible enhancement
due to phase coherence in {\bf II} applies to periodic orbits in either the $E$-shell, or the $E'$-shell.

Semiclassical theory usually restricts itself to integrable systems. This, of course includes the simple case of a single degree of freedom,
for which analogous results to those presented here were obtained by Jarzinsky et all \cite{Jarz}. The obvious generalization would then be to transition probabilities between pairs of pure eigenstates of integrable systems. This strong restriction is here avoided by the averaging 
over a classically narrow energy window, though containing many energy states, which allows the classical evolution to be chaotic, 
or even a generic mixture of regular and irregular motion \cite{Gutzbook}. 
All the same, it may be interesting to compare the present approach to averages of
transitions between individual eigenstates for integrable systems with more than one degree of freedom. 

Most treatments of energy transitions have been carried out within quantum perturbation theory. Thus, it is a great advantage
of the present semiclassical approach that the departure from the initial Hamiltonan is in no way limited. Nonetheless, in the future it may be interesting to make comparisons within the perturbative regime, though this may involve a further development within classical perturbation theory.
Photoionization, as in the historic experiment of Bayfield and Koch \cite{BayKoch}, pushes the target energy level $E'$ into the continuum, 
another extension for the future. It should be noted that, in that case, the initial Hamiltonian $H(\x)$ was the completely 
integrable hydrogen atom, so that it is only its combination with the oscillatory field in $\tilde{\Lambda}(\tau)$ that is chaotic. 

The semiclassical investigation of densities of energy transitions was originally conceived as a new dynamical probe into the nature of the eigenstates of an intrinsically chaotic system, but the explicit incorporation of a general driving Hamiltonian allows for a marked increase of scope. One now realizes that the interplay of the motion generated by both Hamiltonians is the crucial element of the semiclassical theory. 
The seeding of infinitesimal closed compound orbits is remarkable and further insights are presented in Appendix C. Notwithstanding the unfamiliarity of the framework for a full semiclassical calculation of energy transition densities, it is to be hoped that the insight into the fascinating phase space geometry, that is here revealed, will ultimately stimulate realistic computations in the nonperturbative regime.

\appendix

\section{Variational principle for compound trajectories}

It is well established that the time derivative of the action for a trajectory generated by a time-independent Hamiltonian $H(\x)$
is just the constant energy $E$ of the trajectory. This action, may be defined in different ways, the most common being with fixed
end positions, 
\be
\S(\vecq, \vecq'|t) = \int_{\vecq}^{\vecq'} \vecp(\vecq'') \cdot d\vecq'' - E ~t \ ~,
\ee
where $E=H(\vecq, \vecp(\vecq)) = H(\vecq', \vecp(\vecq'))$, so that the canonical transformation 
$\x \mapsto \x'$ is given implicitly by 
\be
\frac{\der \S}{\der \vecq} = \vecp ~~~~  {\rm and} ~~~~ 
\frac{\der \S}{\der \vecq'} = -\vecp' ~.
\ee
The variational principle establishes that this is stationary to first order, with respect to variations of the path 
between the fixed endpoints, $\vecq$ and $\vecq'$. Among these paths, the neighbouring trajectory between the same endpoints is included, 
but with the period $t+\delta t$, so that its energy is $E+\delta E$. Rescaling the time for this trajectory to be just $t$, 
then the  first order difference in action reduces to $-E\delta t$ . 

 A periodic trajectory of period $t$ is also a variational solution, for the choice of endpoints such that $\vecq'=\vecq=\mathbf Q$, and there is again a neighbouring periodic trajectory for the period $t+\delta t$. Then $\delta\S$ for this trajectory can be assessed to first order by comparing
it with a path which slides $\delta \vecp(\delta t)$ instantaneously down to the original $\vecp(\mathbf Q)$, 
completes the original trajectory in the time $t$ and then
slides back to $\vecp + \delta \vecp$ in the time $\delta t$, as shown in Fig A1. Since the full action is stationary, the only
increment comes from this last slide, which, to first order is just $-E\delta t$.

\begin{figure}
\centering
\includegraphics[width=.5\linewidth]{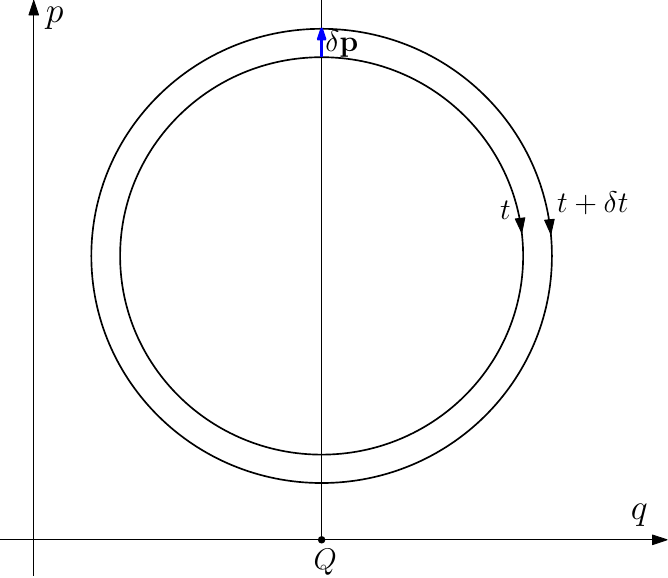}
\caption{The periodic trajectory with period $t$ can be joined to the periodic trajectory with period $t+\delta t$ 
by a vector $(0,\delta \vecp(\delta t))$, in the time $\delta t$. According to the variational principle for trajectories 
with initial and final positions $\vecq'=\vecq=\mathbf Q$, this piecewise smooth path has the same action as the the periodic trajectory with $t+\delta t$, to first order in $\delta t$. Hence the first order change in the action $\delta \S$ is just $-E ~\delta t$.}
\end{figure}

Now let us consider a closed compound trajectory, as described in section 5, with a total period $\tau+t'-\tau-t$.
It is generated by the total time-dependent Hamiltonian: alternating the constant $H(\x)$ (for the intervals $t'$ and $-t$) with
the intrinsically time-dependent $\Lambda(\x, .)$ (for the intervals $\tau$ and $-\tau$). Since this trajectory is uniquely related
to the fixed point of a (compound) canonical transformation, there exists a neighbouring closed compound trajectory
for any alteration of the total time and hence for any alteration to $t+\delta t$ in particular. But then, if one chooses
the position $\vecq = \vecq'= \mathbf Q$ in the range of the trajectory segment with energy $E$, the very same procedure, which established
that the change $\delta \S = -E\delta t$ for a periodic trajectory leads again to the same result, 
regardless of the other segments, since all that is needed is that
the pair of compound trajectories keep close, throughout the total period. This then leads to the simple time derivatives \eref{stat}.

\section{Action for a driven trajectory segment}

A consequence of the Poincar\'e-Cartan theorem is that the invariant \eref{PCinvariant} is null for any closed reducible circuit
on a continuous family of trajectories in the phase space extended by time. Consider then a segment on the the initial shell $H(\x)=E$
at $\tau=0$. This can be pictured as $\tilde{\x}_4(-t)$, that is, the segment in Fig 1b.
The surface patch swept by the trajectories initiated on this segment, which are generated by the Hamiltonian $\Lambda(\x|.))$ 
in the time $\tau$ is limited by the pair of trajectories $\tilde{\x}_1(\tau)$ and $\tilde{\x}_3(\tau)$ (again, 
the latter in the opposite direction of Fig 1b) and by the segment $\tilde{\x}_4(-t|\tau)$ contained in the driven image of the $H(\x)=E$ shell, 
which intersects the $H(\x)=E'$ shell in fig 1a. 

Thus, the Poincar\'e invariant is null over the sequence of four segments that are similar to the compound trajectory, 
except that it is closed by $\tilde{\x}_4(-t|\tau)$ instead of $\tilde{\x}_2$. So the full action for the unknown driven trajectory is 
\begin{eqnarray}
\fl \int_{\tilde{\x}_4(-t|\tau)}[ \vecp \cdot d\vecq - \Lambda(\x|\tau') d\tau']
 = -\int_{\tilde{\x}_1}[ \vecp \cdot d\vecq - \Lambda(\x|\tau') d\tau']
\\  \nonumber
 -\int_{\tilde{\x}_4}[ \vecp \cdot d\vecq - \Lambda(\x|\tau') d\tau'] - \int_{\tilde{\x}_3}[ \vecp \cdot d\vecq - \Lambda(\x|\tau') d\tau'] ~,  
\end{eqnarray} 
as shown in Fig B1. Since the segments $\tilde{\x}_4(-t|\tau)$ and $\tilde{\x}_4$ are traversed in constant $\tau$, this reduces to 
\begin{eqnarray}
\fl \int_{\tilde{\x}_4(-t|\tau)} \vecp \cdot d\vecq  = -\int_{\tilde{\x}_1}[ \vecp \cdot d\vecq - \Lambda(\x|\tau') d\tau']
- \int_{\tilde{\x}_4} \vecp \cdot d\vecq - \int_{\tilde{\x}_3}[ \vecp \cdot d\vecq - \Lambda(\x|\tau') d\tau'] ~. 
\label{drivact} 
\end{eqnarray} 
This is the expression of the energy-action of the orbit segment in the driven shell $H(\x|\tau)=E$, 
which combined with the orbit segment in the $H(\x)=E'$ shell to form the previous bipartite compound orbit in {\bf II}.
 
\begin{figure}
\centering
\includegraphics[width=.5\linewidth]{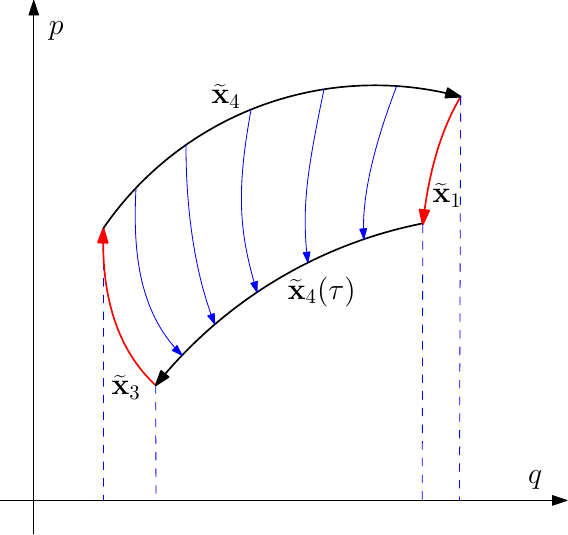}
\caption{The compound trajectory is a reducible circuit on a sheath of trajectories generated by the driving Hamiltonian $\Lambda(\x|\tau')$, starting on the segment $\tilde{\x}_4$. The action for the null Poincar\'e invariant \eref{PCinvariant} may then be considered as a sum of actions for each segment, so that the action for the driven segment $\tilde{\x}_4(-t|\tau)$ ( $\tilde{\x}_4(\tau)$ for short) is given by equation \eref{drivact}. (Note that the actions for $\tilde{\x}_4$ and $\tilde{\x}_1$ are positive, while those for $\tilde{\x}_4(-t|\tau)$ and $\tilde{\x}_3$ are negative.)}
\end{figure}

\section{Commutation of local Hamiltonians}

Let us define local quadratic Hamiltonians, which approximate both our full Hamiltonians in the neighbourhood of the seed $\x_0$ of a CCO family:
\begin{eqnarray} 
H_0(\x)\equiv H(\x_0) +\frac{\der H}{\der \x}|_{\x_0} \cdot (\x-\x_0)
+ \frac{1}{2} (\x-\x_0) \cdot \frac{\der^2 H}{\der \x^2} (\x-\x_0)
\end{eqnarray}  
and
\begin{eqnarray} 
\Lambda_0(\x)\equiv \Lambda(\x_0) 
+\frac{\der \Lambda}{\der \x}|_{\x_0} \cdot (\x-\x_0)
+ \frac{1}{2} (\x-\x_0) \cdot \frac{\der^2 \Lambda}{\der \x^2} (\x-\x_0) ~,
\end{eqnarray}
with an equivalent definition for an initial quadratic time dependent Hamiltonian $\Lambda_0(\x|0)$. Local Hamiltonians have a long history within SC approximations, starting with Heller's {\it frozen Gaussian approximation} \cite{Heller81} and even recently imaginary time propagation for the SC canonical ensemble \cite{deOlOA}.

By construction, the Poisson bracket $\{H_0(\x), \Lambda_0(\x|0)\}=0$ for all $\x$, so that the trajectories generated by $\Lambda(\x|0)$ preserve $H_0(\x)$ (and vice versa). They also commute if starting from the same initial point, that is, one can interchange the finite times $t$ and $\tau$ of the segments and arrive at the same final point. In other words, any such sequential pair of trajectory segments is part of a CCO.

One can also define the quantum Hamiltonians, $\hat{H}_0$ and $\hat{\Lambda}_0$ as appropriately symmetrized variations of $H_0(\hat{p},\hat{q})$ and $\Lambda(\hat{p},\hat{q}|0)$, so that the correspondence principle implies that the quantum commutator also satisfies $[\hat{H}_0, \hat{\Lambda}_0] = 0$.
Therefore, there is no initial energy transition at all between any of the energy levels of $\hat{H}_0$. In contrast, for the full Hamiltonians,  $[\hat{H}, \hat{\Lambda}] \neq 0$, just as $\{H(\x),\Lambda(\x|0)\}\neq 0$ in general. Therefore, the compound trajectory from an arbitrary initial point formed by interchanged segments generated by both Hamiltonians cannot be expected to close in on itself.

What is now perceived is that the local commutation which gives rise to a differential CCO at $\x_0$, shared by both the local and the full Hamiltonians,
is extended, for continuous growth of the times $\tau$, $t$ or $t'$, to a family of commuting trajectory segments, identified with the family of CCO's. 
\footnote{One must distinguish the present local commutation, such that the Poisson bracket cancels at a single point, to a 'local' cancellation of the Poisson bracket at an entire manifold as in \cite{E-RA}. }
This is a three parameter family and, hence, the initial points of its CCO's form a 3-D manifold within the $2N$-D phase space, but within this, the usual role of a commutation that impedes energy transitions, is replaced by quite the opposite: it is the SC support for an oscillatory term of the energy transition density. Each CCO in the continuous family has a segment with energy $H(\tilde{\x(t)})=E$ and $H(\tilde{\x(t')})=E'$, which affects the transition $E\rightarrow E'$.

\section*{Acknowledgments}
I gratefully acknowledge the support of the Pascal Institute of the University of Paris in the program
`Dynamical Foundations of Many-body Quantum Chaos'. Fruitful discussions there with Olivier Brodier and Rodolfo Jalabert
were strong incentives for the extension of the semiclassical transition density to arbitrary Hamiltonians. 
I thank Michael Berry for reminding me of the early example of dynamical quantum chaos and Gabriel Lando for his help in preparing the figures.
Partial financial support from the 
National Institute for Science and Technology - Quantum Information
and CNPq (Brazilian agencies) is gratefully acknowledged.


\section*{Bibliography}
\begin{thebibliography}{99}



\bibitem{Ber89} M.~V. Berry 1989 
         \emph{Proc. R. Soc. Lond. A} {\bf 423} 219
\bibitem{transI} Ozorio de Almeida A.~M. 2022
         \emph{J. Phys. A} {\bf 55} 404004  
\bibitem{transII} Ozorio de Almeida A.~M. 2022
         \emph{J. Phys. A} {\bf 55} 404007 
\bibitem{transIII} Ozorio de Almeida A.~M. 2022
         \emph{Quantum Rep.} {\bf 4} 558-565  	
\bibitem{Jarz} Jarzinsky C, Quan H ~T and Rahav S (2015)
         \emph{Phys. Rev. X} {\bf 5} 031038
\bibitem{OzBro16} Ozorio de Almeida A~M  and Brodier O. 2016
        \emph{J. Phys. A} {\bf 49} 185320
\bibitem{Groen} Groenewold H~J 1946
        \emph{Physica} {\bf 12} 405-460
\bibitem{Moyal} Moyal J~E 1949
        \emph{Proc. Cambridge Phil. Soc.} {\bf 45} 99 
\bibitem{Report} Ozorio de Almeida A.~M. 1998 
        \emph{Phys. Rep.} {\bf 295} 265										
\bibitem{Wigner} Wigner E~P 1932 
         \emph{Phys. Rev.} {\bf 40} 749	
\bibitem{Ber89b} M.~V. Berry 1989 
         \emph{Proc. R. Soc. Lond. A} {\bf 424} 279
\bibitem{SarOA16} Saraceno M and Ozorio de Almeida A~M 2016
				\emph{J. Phys. A} {\bf 49} 185302	
\bibitem{Gutzbook} M.~C. Gutzwiller 1990
        \textsl{Chaos in Classical and Quantum Mechanics} (Springer, New York)
\bibitem{Arnold} Arnold VI  1978 
        \textsl{Mathematical Methods of Classical Mechanics} (Springer, Berlin)	
\bibitem{livro} Ozorio de Almeida A~M 1988 
        \textsl{Hamiltonian Systems: Chaos and Quantization} (Cambridge: Cambridge University Press)
\bibitem{Bar} Baranger M and Davies K~T~R 1987
        \emph{Ann. Phy. N.Y.} {\bf 177} 330-358
\bibitem{Aguiar} de Aguiar M~A~M, Malta C~P, Baranger M and Davies K~T~R 1987
        \emph{Ann. Phy. N.Y.} {\bf 180} 167-205					
\bibitem{BayKoch} Bayfield J. E. and Koch P.~M. 1974
         \emph{Phys.Rev.Lett.} {\bf 33} 298
\bibitem{Heller81} Heller E J 1981
         \emph{J. Chem. Phys.} {\bf 75} 2923–2931
\bibitem{deOlOA}De Oliveira M~G and Ozorio de Almeida A~M 2023
         \emph{J. Stat. Phys.} {\bf 190} 153
\bibitem{E-RA} Escobar-Ruiz A~M and Azuaje R 2024        
 	\emph{J.Phys. A} {\bf 57} 105202		
					



						





 							

 	

 
          
\end{thebibliography}
\end{document}